\documentclass[a4paper,12pt,reqno]{amsart}
\usepackage{amssymb}
\usepackage{graphicx}
\usepackage{natbib}
\calclayout
\theoremstyle{plain}
\newtheorem{Theorem}{Theorem}

\newtheorem{Corollary}[Theorem]{Corollary}
\theoremstyle{definition}
\newtheorem*{Definition}{Definition}
\theoremstyle{remark}
\newtheorem*{Remark}{Remark}
\newcommand{\R}{\mathbb{R}}
\let\cite\citep
\let\mynewpage\relax
\begin{document}

\title[QCC method for reconstructing phylogenetic trees]%
  {Quartet consistency count method for reconstructing phylogenetic trees}

\author{Jin-Hwan Cho}
\address{Department of Mathematics, The University of Suwon}
\email{chofchof@suwon.ac.kr}

\author{Dosang Joe}
\address{Department of Mathematics Education, Konkuk University}
\email{dosjoe@konkuk.ac.kr}

\author{Young Rock Kim}
\address{Department of Mathematics, Konkuk University}
\email{rocky777@math.snu.ac.kr}




\begin{abstract}
Among the distance based algorithms in phylogenetic tree reconstruction,
the neighbor-joining algorithm has been a widely used and effective method.
We propose a new algorithm which counts the number of consistent
quartets for cherry picking with tie breaking.
We show that the success rate of the new algorithm is almost equal
to that of neighbor-joining.
This gives an explanation of the qualitative nature of neighbor-joining
and that of dissimilarity maps from DNA sequence data. 
Moreover, the new algorithm always reconstructs
 correct trees from quartet consistent dissimilarity maps.

\end{abstract}

\maketitle
\mynewpage\section{Introduction}

The neighbor-joining algorithm is widely used among all distance based
methods for phylogenetic tree reconstruction.
In spite of its simplicity neighbor-joining has become a de facto standard
and continued to surface as an effective candidate method for constructing
large phylogenies.
There have been many studies relating to neighbor-joining in many
aspects~\cite{Atteson99, Bryant05, LevyYoshidaPachter06, MihaescuLevyPachter06}.
Questions like how, when, and why neighbor-joining works, have been the
main issues in the empirical and theoretical studies of phylogenetic tree
constructions.

We propose a new algorithm, Quartet Consistency Count abbreviated to QCC,
which gives a partial answer for these questions.
How does the QCC algorithm work?
The QCC algorithm replaces the cherry picking criterion
in neighbor-joining with a new one, the $QC$-criterion in
Theorem~\ref{Theorem 3},
which is to find a pair having maximum quartet consistency counts.
 
The observation is that there are many irrelevant pairwise distances
estimated from DNA sequence data which might reconstruct wrong trees.
The noises or errors from a dissimilarity map are accumulated to pick
irrelevant cherries in neighbor-joining.
However quartet consistency determines how four species are partitioned
into two pairs, and its structure is well preserved in the empirical
DNA sequence data.
It is reasonable to consider quartet consistency rather than adding
the lengths of related edges as neighbor-joining.
 
When does the $QC$-criterion always reconstruct a correct tree?
Atteson proved in~\cite{Atteson99} that neighbor-joining always reconstructs
a correct tree when $l_{\infty}$ radius is $\frac{1}{2}$.
The $QC$-criterion also has the same $l_{\infty}$ radius
which is proved in Corollary~\ref{Corollary 7}.
Unfortunately, very small percentage of DNA sequence data does satisfy
the $l_{\infty}$ radius condition.
However the $QC$-criterion always works under the condition when all quartets
are consistent, which is proved in Theorem~\ref{Theorem 6}.
It is estimated that the quartet consistency rate is relatively high and
strongly related with the success rate of neighbor-joining.

The success rate of QCC is remarkably similar to that of
neighbor-joining even though the tree topologies they generate are
quite different (see Figure~\ref{figure2}).
Nevertheless QCC takes a quite different path in constructing trees
compared to neighbor-joining.
A sample data analysis in Figure~\ref{figure3} shows that the rate of picking
identical cherries in order is less than 65\% even though the two
algorithms generate the same tree topologies.

Why do neighbor-joining and QCC work?
This question is hard to answer.
On the other hand we have seen that the success rates of neighbor-joining
and QCC are almost same.
Since the success of QCC is due to quartet consistency, it is
reasonable to say that neighbor-joining reflects the quartet structure well.
The QCC algorithm gives an explanation of the qualitative nature of neighbor-joining
and that of dissimilarity maps from DNA sequence data. 
\mynewpage\section{Quartet consistency and the $QC$-criterion}

Recall that a dissimilarity map on $[n]:=\{1,2,\dotsc,n\}$ is a function
$d\colon[n]\times[n]\to\R$ such that $d(i,i)=0$ and $d(i,j)=d(j,i)\geq 0$.
A dissimilarity map $d$ is called a \emph{metric} on $[n]$ if the triangle
inequality holds: $d(i,j)\leq d(i,k)+d(k,j)$ for all $i,j,k\in [n]$.
A metric $d$ is a \emph{tree metric} if there exists a tree $T$
with $n$ leaves, labeled by $[n]$, and a non-negative length
for each edge of $T$, such that the length of the unique path
from leaf $x$ to leaf $y$ equals $d(x,y)$ for all $x,y\in [n]$. 
We sometimes write $d_T$ for the tree metric $d$ which is derived
from the tree $T$.

Given four leaves $i,j,k,l$ in a tree $T$, we say that $(ij;kl)$ is
a \emph{quartet} if the path from $i$ to $j$ has no common edge to
the path from $k$ to $l$. In terms of the tree metric $d_T$, it is
equivalent to the following four point condition~\cite{Buneman71}:
\begin{equation} \label{eq:four point condition}
d_T(i,j)+d_T(k,l) \leq d_T(i,k)+d_T(j,l) = d_T(i,l)+d_T(j,k).
\end{equation}

We define a \emph{cherry} of a tree by a pair of leaves
which are both adjacent to the same (internal) node.
This definition of cherry can be reinterpreted as follows:
The pair $\{i,j\}$ is a cherry if and only if $(ij;kl)$ is a quartet
for any pair of leaves $\{k,l\}\subset [n]\setminus\{i,j\}$.
In other words, a cherry of a tree is a pair of leaves
which defines maximum quartets combining with all other pairs,
the number is always $\binom{n-2}{2}$.

Let $d$ be a dissimilarity map on $[n]$. For any $i,j,k,l\in [n]$ we set
\[
w(ij;kl) := \tfrac{1}{4}\bigl[d(i,k)+d(j,l)+d(i,l)+d(j,k)-2[d(i,j)+d(k,l)]\bigr].
\]
In particular, the function $w$ provides a natural weight for quartets,
when $d$ is a tree metric, that is, the length of the path
which connects the path between $i$ and $j$ with the path between $k$ and $l$.

The neighbor-joining algorithm makes use of the following cherry picking
theorem~\cite{SaitouNei87} by peeling off cherries
to recursively build a tree. 

\begin{Theorem} \label{Theorem 1}
If $d$ is a tree metric on $[n]$, then any pair of leaves that maximizes
$Z_d(i,j)=\sum_{\{k,l\}\subset [n]\setminus\{i,j\}} w(ij;kl)$
is a cherry in the tree.
\end{Theorem}

An equivalent, but computationally superior, formulation is
the following $Q$-criterion~\cite{StudierKeppler88},
which is the unique selection criterion in some sense \cite{Bryant05}.

\begin{Corollary} \label{Corollary 2}
If $d$ is a tree metric on $[n]$, then any pair of leaves that minimizes
$Q_d(i,j)=(n-2)d(i,j)-\sum_{k\neq i} d(i,k)-\sum_{k\neq j} d(j,k)$
is a cherry in the tree. 
\end{Corollary}

We now introduce the notion of quartet consistency and then propose
a new criterion called the $QC$-criterion which counts the number of
consistent quartets to determine the cherries.

\begin{Definition}
A dissimilarity map $d$ is \emph{quartet consistent} with a tree $T$ if
\begin{equation} \label{eq:quartet consistency condition}
d(i,j)+d(k,l) \leq \min\{d(i,k)+d(j,l), d(i,l)+d(j,k)\}
\end{equation}
for all quartets $(ij;kl)$ in $T$. Note that any tree metric $d_T$ is
quartet consistent with $T$ since $d_T$ satisfies the four point
condition~(\ref{eq:four point condition}).
\end{Definition}

\begin{Remark}
In terms of the weight function $w$, the quartet consistency condition
(\ref{eq:quartet consistency condition}) is equivalent to
$w(ij;kl) \geq \max\{w(ik;jl),w(il;jk)\}$ which is used
in~\cite[Definition~8]{MihaescuLevyPachter06}.
\end{Remark}

\begin{Theorem} \label{Theorem 3}
If $d$ is a tree metric on $[n]$, then any pair of leaves that maximizes
\begin{align*}
QC_d(i, j) &:= \text{\upshape the number of pairs $\{k,l\}\subset[n]\setminus\{i,j\}$ such that} \\
&\phantom{:=}\qquad d(i,j)+d(k,l) \leq \min\{d(i,k)+d(j,l), d(i,l)+d(j,k)\}
\end{align*}
is a cherry in the tree.
\end{Theorem} 

\begin{proof}
Since $d$ is a tree metric, the four point
condition~(\ref{eq:four point condition}) implies that
$QC_d(i,j)$ equals the number of pairs $\{k,l\}\subset[n]\setminus\{i,j\}$ such that $(ij;kl)$ is a quartet, which becomes the maximum number $\binom{n-2}{2}$ if and only if $\{i,j\}$ is a cherry.
\end{proof}

The following theorem has been a widely used justification
for the observed success of neighbor-joining.

\begin{Theorem}[Atteson~\cite{Atteson99}] \label{Theorem 4}
Neighbor-joining has $l_{\infty}$ radius $\frac{1}{2}$.
\end{Theorem}

This implies that neighbor-joining always reconstruct a correct tree
if the distance estimates are at most half the minimal edge length
of the tree away from their true value.
Two conditions are introduced in~\cite{MihaescuLevyPachter06}
to explain why neighbor-joining is useful in practice. One is
quartet consistent and the other is quartet additive
which appears to be rather technical.
It is also verified that Atteson's theorem is a special case of
the following theorem~\cite[Theorem 17]{MihaescuLevyPachter06}.

\begin{Theorem} \label{Theorem 5}
If $d$ is quartet consistent and quartet additive with a tree $T$, then neighbor-joining applied to $d$ will construct a tree with same topology as $T$.
\end{Theorem}

Atteson's condition is sufficient to satisfy
the quartet consistent and quartet additive condistions.  
Since these two conditions are not always satisfied, the success rate
of reconstructing a correct tree by neighbor-joining is limited.
In practical computation, however, the pairwise distances are estimated
from noisy data, and consequently, the resulting dissimilarity map is very
unlikely to be a tree metric.
The dissimilarity map by estimating distances from DNA sequence data
does not satisfy the quartet consistency and quartet additive conditions
in most cases even when neighbor-joining is successful.
In practical sense, it is not fully understood why neighbor-joining
is successful.

We state the consistency theorem for the $QC$-criterion. It says that
the $QC$-criterion for cherry picking with the same reduction step as
neighbor-joining always reconstruct a correct tree whenever a dissimilarity
map is quartet consistent.

\begin{Theorem} \label{Theorem 6}
If a dissimilarity map $d$ is quartet consistent with a tree $T$,
then the $QC$-criterion for cherry picking with the reduction step
of neighbor-joining applied to $d$ will construct a tree with
the same topology as $T$.
\end{Theorem}

\begin{proof}
Since $d$ is quartet consistent with $T$, $QC_d(i,j)$ is greater
or equal to the number of pairs $\{k,l\}\subset[n]\setminus\{i,j\}$
such that $(ij;kl)$ is a quartet,
which becomes the maximum number $\binom{n-2}{2}$ when $\{i,j\}$ is a cherry
in $T$.
Therefore, the $QC$-criterion always picks a cherry
if $d$ is quartet consistent with $T$.
It suffices to show that the quartet consistency condition is preserved
in the reduction step of neighbor-joining.

Suppose that $\{i,j\}$ is a cherry picked in the previous step.
The reduction step of neighbor-joining constructs
the reduced tree $\widetilde T$ by removing
the two leaves $i,j$ and adding a new one $i_*$.
The dissimilarity map is also modified by the equation
$d(i_*,k)=\frac{1}{2}\bigl[d(i,k)+d(j,k)-d(i,j)\bigr]$
for all $k\in[n]\setminus\{i,j\}$.
We will show that the modified dissimilarity map is quartet consistent with $\widetilde T$. Note that $(i_*k;lm)$ is a quartet in $\widetilde T$ if and only if $(ik;lm)$ and $(jk;lm)$ are both quartets in $T$.

Suppose $(i_*k;lm)$ is a quartet in $\widetilde T$, then we have  
\begin{gather*}
d(i,k)+d(l, m) \leq \min \bigl\{ d(i, l)+d(k,m), d(i,m)+d(k, l)\bigr\}, \\
d(j,k)+d(l, m) \leq \min \bigl\{ d(j, l)+d(k,m), d(j,m)+d(k, l)\bigr\},
\end{gather*}
since $d$ is quartet consistent with $T$.
Combining these two inequalities, we get
\begin{align*}
& d(i,k)+d(j,k)+2d(l, m) \\
&\qquad \leq\min\{d(i, l)+d(j, l)+2d(k,m), d(i, m)+d(j, m)+2d(k,l)\}.
\end{align*}
Therefore
\begin{align*}
& d(i_*,k)+d(l,m) = 
\tfrac{1}{2}\bigl[d(i,k)+d(j,k)+2d(l,m)-d(i,j)\bigr] \\
&\quad\leq \min\bigl\{ \tfrac{1}{2}[d(i,l)+d(j,l)-d(i,j)]+d(k,m), \tfrac{1}{2}[d(i,m)+d(j,m)-d(i,j)]+d(k,l)\bigr\}\\
&\quad= \min\{ d(i_*,l)+d(k,m),d(i_*,m)+d(k,l)\}. \qedhere
\end{align*}
\end{proof}

We can also prove that the $QC$-criterion has $l_\infty$ radius $\frac{1}{2}$. This means, like neighbor-joining, if the distance estimates are at most half the minimal edge length of the tree away from their true values then the $QC$-criterion will reconstruct a correct tree.
It was proved in \cite[Corollary 20]{MihaescuLevyPachter06} that the $l_\infty$ radius
$\frac{1}{2}$ condition implies the quartet consistent and quartet additive conditions. We would like to include a short proof of it to make this paper self-contained.

\begin{Corollary} \label{Corollary 7}
The $QC$-criterion has $l_{\infty}$ radius $\frac {1}{2}$.
\end{Corollary}

\begin{proof}
Suppose that distance estimates are at most half of the minimal edge length of the tree. Then it is quartet consistent with it.
Since $\min\bigl\{d(i,k)+d(j,l),d(i,l)+d(j,k)\bigr\}-\bigl[d(i,j)+d(k,l)\bigr]$ is less than four times of maximum noises minus two times of length of connecting edge associated with the quartet $(ij,kl)$, if maximum error is less than half of the minimal edge length, the quartet structure is consistent with the tree.
\end{proof}

Unlike neighbor-joining, the selection criterion $QC$ is not distance linear~\cite{Bryant05}. It rather depends on how a dissimilarity map preserves the quartet structures of a given tree.

\begin{Remark}
In \cite[Example~11]{MihaescuLevyPachter06}, they constructed a quartet consistent metric on an eight leaves tree which cannot be reconstructed by neighbor-joining. By Theorem~\ref{Theorem 6}, $QC$-criterion will reconstruct the correct tree.
\end{Remark}
\mynewpage\section{Performance of the quartet consistency count algorithm}

The Quartet Consistency Count algorithm consists of two steps,
one is the cherry picking step and the other is the reduction step.
It adopts the $QC$-criterion instead of the $Q$-criterion of neighbor-joining for the cherry picking step, but the same algorithm for the reduction step
as neighbor-joining.

We sometimes get different tree topologies for one dissimilarity map if
the $QC$-criterion is used solely in the cherry picking step.
This happens when there are more than one pair having the same quartet consistency count.
In this case the order of picking cherries depends on the order of leaves in
the input data, and the resulting tree might have different topologies.
To overcome the defect a tie-breaking routine is required in the QCC algorithm.

We have tested several tie breaking methods, one of which gives a penalty
for the bad case when the inequality
$d(i,j)+d(k,l)>\max\{d(i,k)+d(j,l),d(i,l)+d(j,k)\}$ happens,
and another one minimizing the sum of errors, $|d(i,k)+d(j,l)-d(i,l)-d(j,k)|$.
Most of all, minimizing the value $Q_{d}(i,j)$ in Corollary~\ref{Corollary 2}
gave a better success rate, and it was adopted for the tie breaking routine
in the QCC algorithm as follows:

{\setlength{\parindent}{0pt}
\bigskip\hrule\smallskip
\textbf{Quartet Consistency Count Algorithm}
\smallskip\hrule\medskip
\textbf{Input:} A dissimilarity map $d$ on the set $[n]$
\par\smallskip
\textbf{Output:} A phylogenetic tree $T$ whose tree metric $d_T$ is close to $d$
\par\smallskip
\textbf{Cherry picking step:}
Find a pair $\{i,j\}$ having the maximum $QC_{d}(i,j)$ count.
If there are more than one such pair, choose a pair having the minimum
$Q_{d}(i,j)$ value among them.
\par\smallskip
\textbf{Reduction step:}
Remove $\{i,j\}$ from the tree, thereby creating a new leaf $i_*$.
For each leaf $k$ among the remaining $n-2$ leaves, set
$d(i_*,k) = \frac{1}{2}[d(i,k)+d(j,k)-d(i,j)]$. 
Return to the cherry picking step until there are no more leaves to collapse. 
\smallskip\hrule\bigskip}

\subsection*{Success rates of QCC and neighbor-joining}
\begin{figure}
\centering
\includegraphics{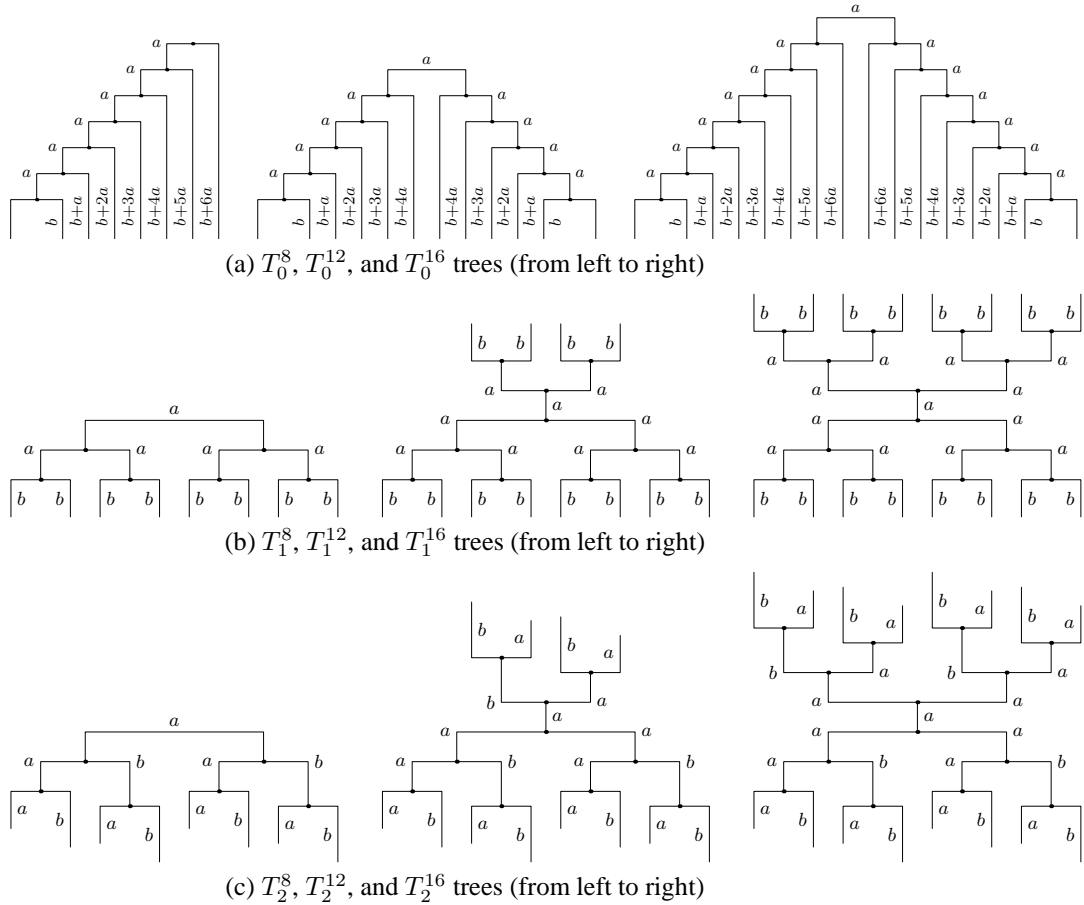}
\caption{Nine tree shapes $T_0^n$, $T_1^n$, and $T_2^n$ for $n=8$, $12$, and $16$}
\label{Figure1}
\end{figure}

The success rate of QCC is discussed in the perspective of neighbor-joining.
We tested QCC with simulated data on the two parameter family of trees
described in~\cite{SaitouNei87}. We simulated 1,000 data sets on each of
the nine tree shapes, $T_0^n$, $T_1^n$, and $T_2^n$ when the number of
leaves $n=8$, $12$, and $16$ (see Figure~\ref{Figure1})
at the three edge length ratios, $a/b=0.01/0.04$, $0.02/0.13$, $0.03/0.34$ for $T_0$, and $a/b=0.01/0.07$, $0.02/0.19$, $0.03/0.42$ for $T_1$ and $T_2$.
This was repeated three times for sequences of length 500, 1000, and 2000 bp.
The Juke-Cantor distance method for GTR model was used to get pairwise
distances from the simulated DNA sequence data generated by \texttt{Seq-Gen}~\cite{RambautGrassly97}.

Tabel~\ref{Table1} shows the success rate of QCC compared with neighbor-joining.
The numbers inside parentheses are the differences between the success
rate of QCC and that of neighbor-joining, positive (resp.~negative) numbers
represent that the success rate of QCC is better (resp.~worse)
than that of neighbor-joining.
It is remarkable that the success rates of the two algorithms are almost same,
and that the differences are independent of the tree shapes and the bp lengths
of simulated DNA sequence data.

\begin{table}
\begingroup
\renewcommand{\baselinestretch}{1.0}
\renewcommand{\arraystretch}{1.2}
\small
\begin{tabular}{|c|ccc|ccc|ccc|} \hline
bp & \multicolumn{3}{c|}{500} &\multicolumn{3}{c|}{1000} &\multicolumn{3}{c|}{2000} \\\hline\hline
\rule[-1.5ex]{0pt}{3ex} $a/b$ 
      & $\frac{0.01}{0.04}$ & $\frac{0.02}{0.13}$ & $\frac{0.03}{0.34}$
      & $\frac{0.01}{0.04}$ & $\frac{0.02}{0.13}$ & $\frac{0.03}{0.34}$
      & $\frac{0.01}{0.04}$ & $\frac{0.02}{0.13}$ & $\frac{0.03}{0.34}$ \\\hline
$T_0^{8}$
& 68.4  & 50.7  & 10.9  & 91.6  & 82.8  & 26.3   & 99.4  & 96.9  & 56.5 \\[-1ex]
& (-0.2) & (-0.3) & (-0.3) & (0.0) & (0.0) & (0.7) & (0.0) & (0.0) & (-0.8) \\
$T_0^{12}$
& 63.7 & 44.5  & \phantom{0}4.2  & 93.7  & 85.0  & 21.0 & 99.9  & 99.0  & 59.1  \\[-1ex]
& (0.1) & (0.1) & (-0.2) & (-0.1) & (-0.7) & (-0.3) & (0.0) & (-0.5) & (-0.3) \\
$T_0^{16}$
& 39.0 & 20.3 & \phantom{0}0.2 & 83.9 & 65.2 & \phantom{0}5.4 & 99.3  & 96.0  & 35.1   \\[-1ex]
& (1.6) & (-0.2) & (-0.1) & (-0.2) & (-0.5) & (0.5) & (0.0) & (-0.9) & (-1.1)\\
\hline\hline
\rule[-1.5ex]{0pt}{3ex} $a/b$ 
      & $\frac{0.01}{0.07}$ & $\frac{0.02}{0.19}$ & $\frac{0.03}{0.42}$
      & $\frac{0.01}{0.07}$ & $\frac{0.02}{0.19}$ & $\frac{0.03}{0.42}$
      & $\frac{0.01}{0.07}$ & $\frac{0.02}{0.19}$ & $\frac{0.03}{0.42}$ \\\hline
$T_1^{8}$
& 72.5  & 55.9  & 10.8  & 95.4  & 86.7 & 32.6 & 99.9  & 98.7  & 65.8 \\[-1ex]
& (0.0) & (-0.3) & (-0.6) & (-0.1) & (-0.2) & (0.1) & (0.0) & (0.0) & (0.1)\\
$T_1^{12}$
& 59.9 & 44.0 & \phantom{0}3.0  & 93.5 & 81.3 & 24.3 & 99.7  & 99.0  & 65.1 \\[-1ex]
& (0.2) & (0.2) & (0.6) & (0.1) & (0.0) & (0.0) & (0.0) & (0.0) & (0.3) \\
$T_1^{16}$
& 51.0 & 32.3 & \phantom{0}1.8 & 92.0  & 80.7 & 15.0 & 99.6 & 98.6 & 55.2 \\[-1ex]
& (0.6)  & (0.3) & (-0.4) & (0.5) & (0.4) & (-0.1) & (0.0) & (-0.1) & (0.9)\\
\hline\hline
\rule[-1.5ex]{0pt}{3ex} $a/b$ 
      & $\frac{0.01}{0.07}$ & $\frac{0.02}{0.19}$ & $\frac{0.03}{0.42}$
      & $\frac{0.01}{0.07}$ & $\frac{0.02}{0.19}$ & $\frac{0.03}{0.42}$
      & $\frac{0.01}{0.07}$ & $\frac{0.02}{0.19}$ & $\frac{0.03}{0.42}$ \\\hline
$T_2^{8}$
& 81.5  & 68.2  & 19.0 & 96.4  & 91.3 & 44.2 & 99.9  & 98.6  & 70.0  \\[-1ex]
& (-0.1) & (0.0) & (0.4) & (0.0) & (-0.1) & (-0.4) & (0.0) & (0.0) & (-0.1) \\$T_2^{12}$
& 69.0  & 55.8  & \phantom{0}4.3  & 96.6  & 89.7 & 26.4 & 99.8  & 99.5  & 60.8 \\[-1ex]
& (-0.5) & (0.4) & (0.3) & (0.0) & (-0.3) & (-1.1) & (0.0) & (0.0) & (0.1) \\$T_2^{16}$
& 64.7  & 47.3  & \phantom{0}2.2  & 95.5  & 87.2 & 17.9 & 99.9 & 99.3 & 61.0 \\[-1ex]
& (0.0) & (-0.2) & (0.0) & (0.0) & (0.3) & (2.5) & (0.0) & (-0.1) & (-0.4) \\\hline
\end{tabular}
\endgroup

\bigskip
\caption{Success rate of QCC compared with neighbor-joining:
The values denote the success rate of neighbor-joining in percentage,
and the numbers inside parentheses represent the difference of success rates
of QCC compared with neighbor-joining.}
\label{Table1}
\end{table}

\begin{figure}
\centering
\includegraphics{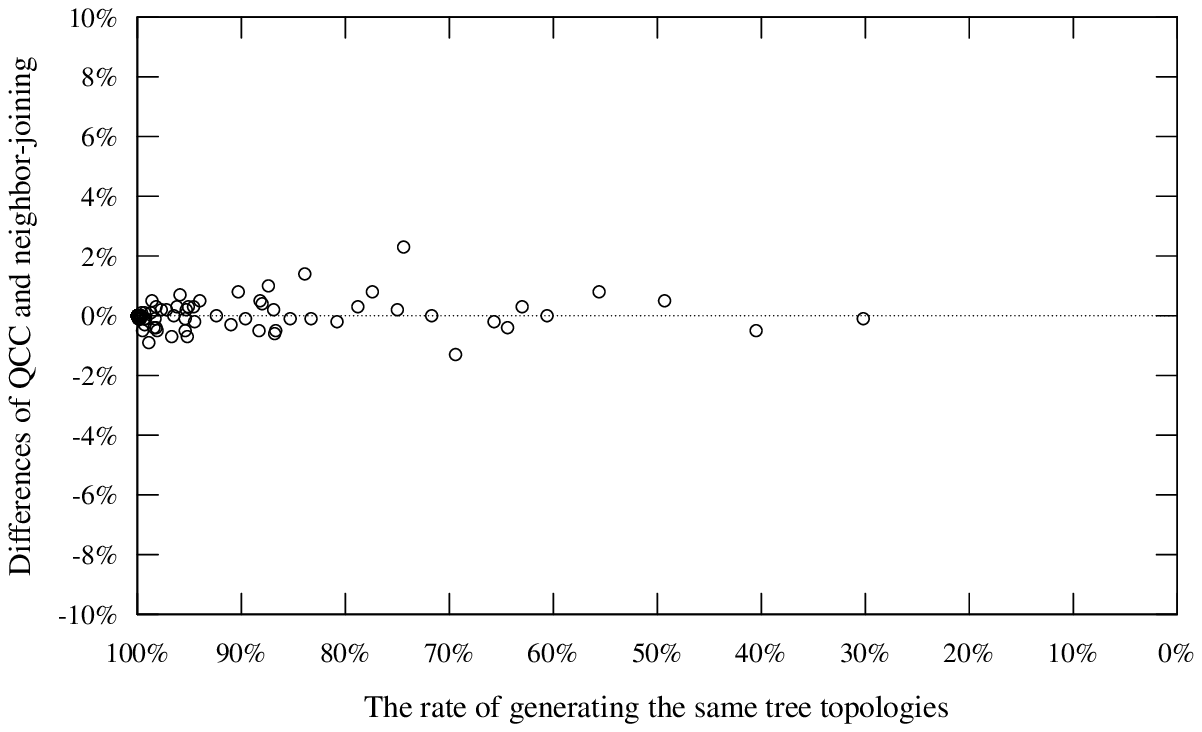}
\caption{Differences of the success rates of neighbor-joining and QCC
according to the rate of generating the same tree topologies}
\label{figure2}
\end{figure}

Figure~\ref{figure2} shows an interesting fact that the differences
do not vary even if the tree topologies generated by the two algorithms
are quite different.
Note that the difference rate is still quite small when the rate of
generating the same tree topologies is around 30\%. 

\subsection*{Independent cherry picking order}
\begin{figure}
\centering
\includegraphics{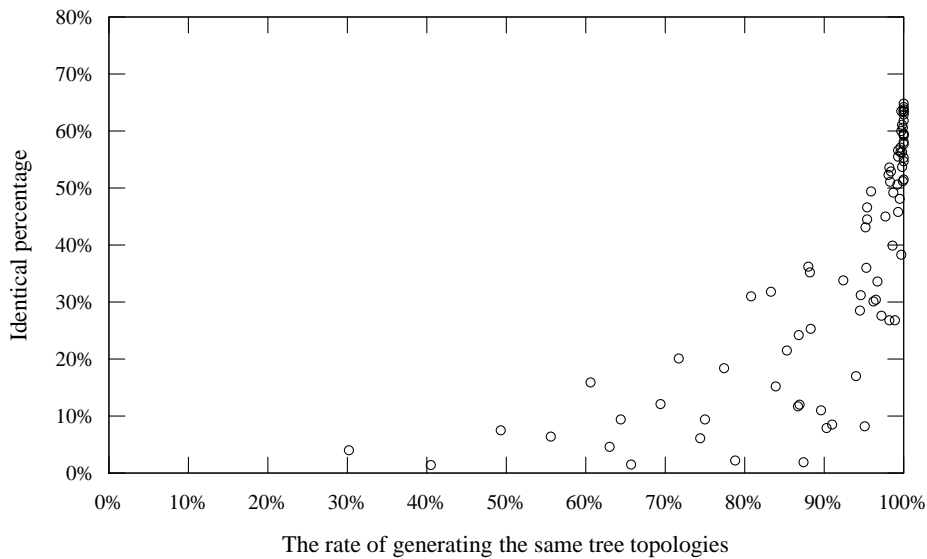}
\caption{The percentage of picking identical cherries in order
according to the rate of generating the same tree topologies}
\label{figure3}
\end{figure}

Even success rates of QCC and neighbor-joining are almost same to each other,
the paths of picking cherries in order are quite different.
We investigated the percentage of picking identical cherries in order out of
1000 data sets for each 81 different trees.
It is interesting to see in Figure~\ref{figure3} that the identical
percentage is not so high even QCC and neighbor-joining generate the
same tree topologies.
When the rate of generating the same tree topologies is more than 95\%,
the identical percentage does not exceed 65\% in the simulated data sets.
It indicates that the QCC algorithm takes quite different paths of
picking cherries compared to neighbor-joining.

\subsection*{Quartet consistency rate and neighbor-joining}
Quartet consistency rate of a dissimilarity map is the percentage of
four leaves satisfying the quartet consistency condition
(\ref{eq:quartet consistency condition})
with a given tree $T$ over all possible quartets in $T$.
The QCC algorithm heavily depends on this rate, for instance,
it recovers a correct tree when the rate is 100\% by Theorem~\ref{Theorem 6}.

\begin{figure}[h]
\centering
\includegraphics{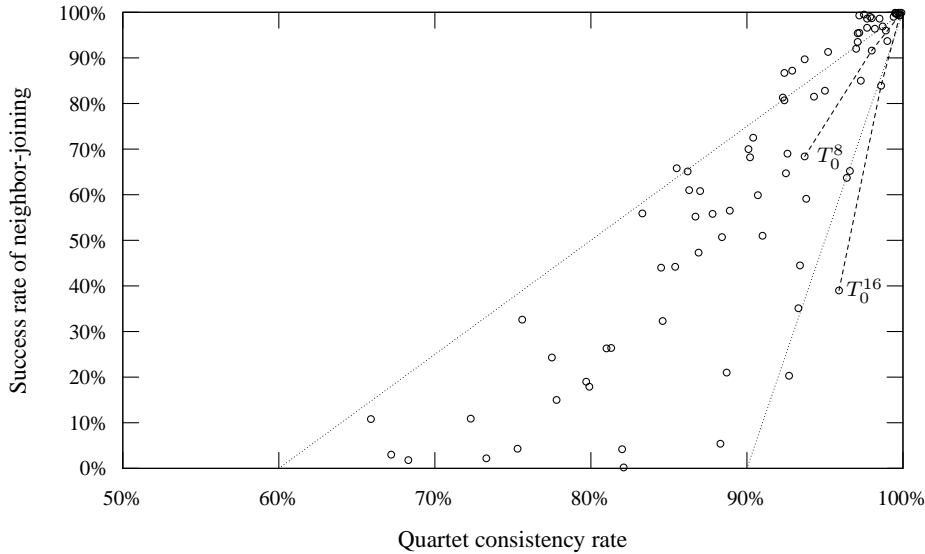}
\caption{Quartet consistency rate with respect to the success rate
of neighbor-joining}
\label{figure4}
\end{figure}

We investigated in Figure~\ref{figure4} that the correlation of
quartet consistency rate with respect to the success rate of neighbor-joining.
The correlation coefficient was computed as 0.8736.
The graph shows that the success rate of neighbor-joining near 100\% is
almost same as quartet consistency, as we expected,
since the success rates of QCC and neighbor-joining are almost same.
Quartet consistency rates also increase as bp lengths increase.
The dashed line in the graph, denoted by $T^{8}_{0}$ (resp.~$T^{16}_{0}$)
connects the three points representing the success rates of neighbor-joining
for the tree $T^{8}_{0}$ (resp.~$T^{16}_{0}$) with the ratio $a/b=0.01/0.04$ when the bp lengths are 500, 1000, and 2000.

\mynewpage\section{Discussion}

\subsection*{Quartet based methods}
There are many quartet based methods in reconstructing the phylogenetic trees.
Several methods were proposed in~\cite{BryantSteel01} to construct the optimal
trees which agree with the largest number of quartets or the maximum weight
set of quartets.
The general problems are known to be NP-hard. The implemented algorithms, Quartet-Cleaning and $Q^*$, have quite different nature
statistically compared to neighbor-joining~\cite{JohnWarnowMoretVawter03}.
The QCC algorithm is quite different to the well-known quartet based methods
derived from quartet puzzling problem, it is shown to be close to
neighbor-joining.

\subsection*{$QC$-criterion without tie-breaking}
\begin{figure}
\centering
\includegraphics{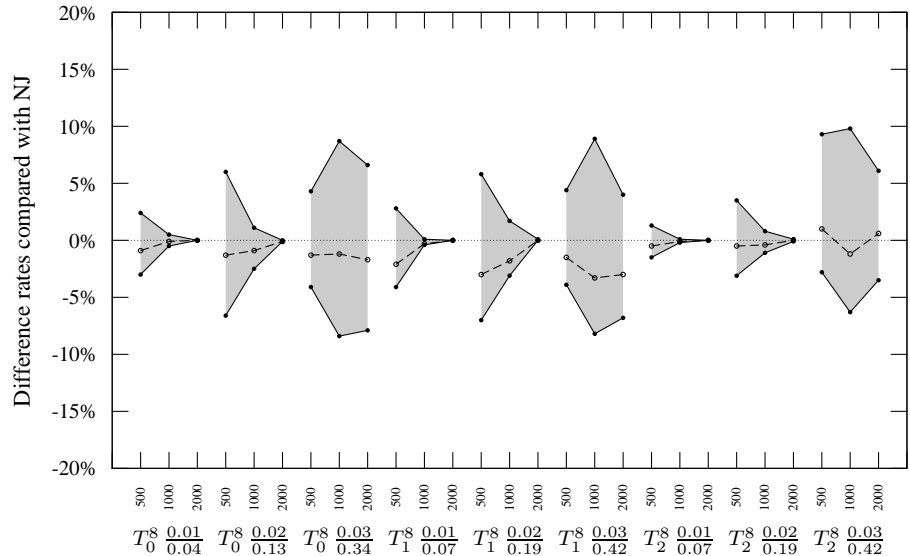}
\caption{$QC$-criterion without tie breaking}
\label{figure5}
\end{figure}

The cherry picking step in the QCC algorithm requires a tie-breaking routine
to avoid the dependency of the order of the leaves in the input data.
To estimate the best and the worst behavior of the algorithm without
tie-breaking, we shuffled the order of the leaves 100 times randomly, and then counted how many correct trees are reconstructed.
By counting as a success when there is at least one such correct
tree out of 100 trials, we get the best success rate.
On the other hand, the worst success rate follows if we count as a success
when the correct tree is always reconstructed for all trials.
The upper and lower solid lines in Figure~\ref{figure5} represent
the best and the worst success rates, respectively.
The dashed line in the middle represents the average of the counts.

As the figure shows, it might be possible to have a good tie-breaking routine
which gives a better success rate than that of neighbor-joining.
We believe that a deeper understanding of tie-breaking routine of the QCC algorithm should have more results in this direction.

\subsection*{Conclusion}
The behavior of the QCC algorithm is similar to that of neighbor-joining.
From this similarity QCC reflects the qualitative nature of neighbor-joining
and that of dissimilarity maps from DNA sequence data.
The QCC algorithm has the same $l_{\infty}$ radius $\frac{1}{2}$ as
neighbor-joining, and it requires only the quartet consistency condition
to reconstruct a correct tree.
\mynewpage\section*{Acknowledgements}
This work was supported by grant No.~R01-2006-000-10047-0 from
the Basic Research Program of the Korean Science \& Engineering
Foundation for second author.
Third author was supported in part by KRF(grant No.~2005-070-C00005
and grant No.~R14-2002-007-01001-0).
\mynewpage

\begin{thebibliography}{}

\bibitem[Atteson, 1999]{Atteson99}
Atteson, K. 1999.
\newblock The performance of {N}eighbor-{J}oining methods of phylogenetic
  reconstruction.
\newblock {\em Algorithmica}, 25(2--3):251--278.

\bibitem[Bryant, 2005]{Bryant05}
Bryant, D. 2005.
\newblock On the uniqueness of the selection criterion in {N}eighbor-joining.
\newblock {\em J. Classification}, 22(1):3--15.

\bibitem[Bryant and Steel, 2001]{BryantSteel01}
Bryant, D. and Steel, M. 2001.
\newblock Constructing optimal trees from quartets.
\newblock {\em J. Algorithms}, 38(1):237--259.

\bibitem[Buneman, 1971]{Buneman71}
Buneman, P. 1971.
\newblock The recovery of trees from measures of dissimilarity.
\newblock In Hodson, F.~R., Kendall, D.~G., and Tautu, P., editors, {\em
  Mathematics in Archeological and Historical Sciences}, pages 387--395.
  Edinburgh University Press.

\bibitem[John et~al., 2003]{JohnWarnowMoretVawter03}
John, K.~S., Warnow, T., Moret, B., and Vawter, L. 2003.
\newblock Performance study of phylogenetic methods: (unweighted) quartet
  methods and neighbor-joining.
\newblock {\em J. Algorithms}, 48(1):173--193.

\bibitem[Levy et~al., 2006]{LevyYoshidaPachter06}
Levy, D., Yoshida, R., and Pachter, L. 2006.
\newblock Beyond pairwise distances: neighbor-joining with phylogenetic
  diversity estimates.
\newblock {\em Mol. Biol. Evol.}, 23(3):491--498.

\bibitem[Mihaescu et~al., 2006]{MihaescuLevyPachter06}
Mihaescu, R., Levy, D., and Pachter, L. 2006.
\newblock Why neighbor-joining works.
\newblock arXiv.org:cs/0602041.

\bibitem[Rambaut and Grassly, 1997]{RambautGrassly97}
Rambaut, A. and Grassly, N. 1997.
\newblock {S}eq-{G}en: {A}n application for the {M}onte {C}arlo simulation of
  {DNA} sequence evolution along phylogenetic trees.
\newblock {\em Comput. Appl. Biosci.}, 13:235--238.

\bibitem[Saitou and Nei, 1987]{SaitouNei87}
Saitou, N. and Nei, M. 1987.
\newblock The neighbor-joining method: {A} new method for reconstructing
  phylogenetic trees.
\newblock {\em Mol. Biol. Evol.}, 4(1):406--425.

\bibitem[Studier and Keppler, 1988]{StudierKeppler88}
Studier, J.~A. and Keppler, K.~J. 1988.
\newblock A note on the neighbor-joining method of {S}aitou and {N}ei.
\newblock {\em Mol. Biol. Evol.}, 5:729--731.

\end{thebibliography}
\end{document}